\documentclass[aps,twocolumn,groupedaddress,showpacs,floats,amssymb]{revtex4}
\usepackage[dvips]{graphicx}

\begin{document}

\noindent {\bf Comment on ``Twofold Advance in the Theoretical Understanding of Far-From-Equilibrium Properties of Interacting Nanostructures''}\vspace{2mm}\\
Boulat, Saleur and Schmitteckert (BSS)\cite{BSS} report
results on the full $I-V$ characteristics of the interacting resonant level 
model (IRLM) exhibiting region with unexpected negative differential
conductance (NDC). Using time-dependent density matrix renormalization group 
complemented with the exact solution performed at a special point (the
self-dual point) in the parameter space BSS have shown that at nonzero
Coulomb interaction $U$ the current flowing through the impurity level (IL) 
exhibits
a power-law asymptotics as a function of large applied bias voltage. Similar 
conclusion was earlier reached by Doyon\cite{Doyon}.
\begin{figure}
\centerline{\includegraphics[width=0.99\columnwidth, clip]{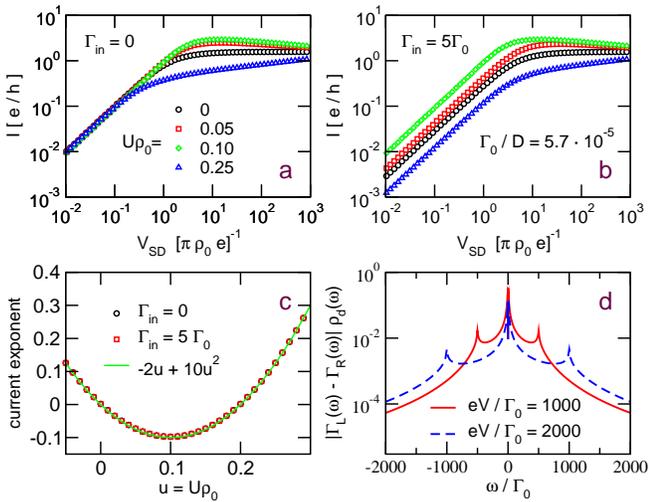}}
\caption{(Color online) 
Panels a-b: The $I-V$ characteristics of the $N=10$ channel IRLM
calculated in the framework of the perturbative approach 
of Ref.\cite{BVZ}, for inelastic rate $\Gamma_{in}=0$ and $5\Gamma_0$,
respectively. For finite Coulomb interaction $U$, $I$
exhibits a power law asymptotics at large $V$. 
The corresponding exponent shown in panel c,  essentially 
independent of $\Gamma_{in}$. Panel d shows 
the mechanism leading to the decay of the current:
spectral weight is transferred outside the voltage window.  
}
\label{Fig1}
\end{figure}
Even though their results are solid and supported by both analytic and numeric
arguments, BSS 
%admit in the conclusion of their paper 
concluded
that 
``the NDC
at large voltage seems a truly nonperturbative behavior,
with unclear physical origin''.
On the contrary, the remarkable physics of 
NDC
can be explained by simple physical arguments and, as we shall show in this
Comment, in certain circumstances can be calculated in the framework of
perturbation theory.

As it was shown by Vlad\'ar and the present authors\cite{BVZ}, 
by increasing the number of screening channels $N$ in the IRLM, the 
turning point of the hybridization exponent (cf. the self-dual point 
for $N=2$) can be pushed down to the perturbative regime. 
%In that paper it was shown that 
The scattering matrix elements 
%can be 
were
evaluated up to order $U^2$, 
%even if
%the conduction electron states obey non-equilibrium distributions. 
in the leading order of RG method.
Using
the scattering matrix elements we can derive rate equations (RE) to determine
the occupancy of the IL
as well as the current $I$ flowing through
it. The RE approach 
takes only the sequential tunneling (ST) through the IL
into account\cite{Datta}
neglecting coherent co-tunneling (CT) processes,
in contrast to the calculation of BSS. Nevertheless, for small 
hybridization $\Gamma$
ST is dominant thus our results are reliable in that
regime. To check that, we have repeated the calculation with the inclusion of
a large inelastic rate $\Gamma_{in}=5\Gamma_0$ to de-phase electrons on the
IL therefore suppressing CT 
%processes 
($\Gamma_0$ being the
bare value of $\Gamma$). In panels a-b of Fig.~\ref{Fig1}, the
$I-V$ curves are shown for $\Gamma_{in}=0$ and $=5\Gamma_0$,
respectively. $I$ clearly exhibits a power-law asymptotics at large
$V$. As shown in panel c, the corresponding exponent practically does not
depend on the value of $\Gamma_{in}$, supporting the validity of 
ST approximation. It is also shown in panel c that the exponent 
of $I$ coincide with the equilibrium 
%renormalization 
exponent of 
%the hybridization 
$\Gamma(\omega)\sim\Gamma_0[\omega / D]^{-2u+Nu^2}$, where $u$
is the dimensionless Coulomb interaction $u=U\varrho_0$, $\varrho_0$ being
the density of states (DoS) of conduction electrons per channel. 
Panel d illustrates the mechanism leading to NDC for $u=0.1$: by
plotting the IL DoS properly weighted with $\Gamma$-s 
(cf. the combination which enters the expression of the
current) it is clear that spectral weight is transformed outside the voltage
window. 
%One can even understand why 
In the ST approximation
the exponent of $I$ coincide
with the equilibrium exponent of $\Gamma$. 
%Given that $V$ is 
For
large enough $V$, the expression of $I$ can be simplified
dramatically, $I\sim\Gamma(eV)=\Gamma_L(eV)+\Gamma_R(eV)$. 
%This result can be
%understood as follows: 
The saturation value of $I$ is given by
$\Gamma$ which gets renormalized in presence of
$U$
but in non-equilibrium situation the flow is terminated at
$\omega\sim eV$, thus $I\sim\Gamma(eV)=\Gamma_0[eV/D]^{-2u+Nu^2}$.
In general, our result does not rely on perturbative arguments: having 
the exact exponent of $\Gamma$ from a reliable equilibrium
calculation (e.g. numerical RG) one
should be able to accurately describe the asymptotics of $I$ 
{\em in the ST regime}.
Note that in the calculation of BSS $\Gamma_0$
was not very small therefore the CT 
%processes were 
was not
negligible. That is why we found only a qualitative
agreement between their exponents of $I$ and those
of $\Gamma$ extracted from NRG.

We conclude that in the ST regime of IRLM the NDC is a result of 
renormalization of $\Gamma$. This research was supported by Hungarian Grants
OTKA No. T048782.  

\noindent L. Borda$^{1,2}$, A. Zawadowski$^{2}$

\noindent
$^{1}$ Physikalisches Institut, Universit\"at Bonn, Germany\\
$^{2}$ Research Group of HAS, TU Budapest, Hungary


\vspace{-0.4cm}

\begin{thebibliography}{50}
\bibitem{BSS} E. Boulat, H. Saleur, and P. Schmitteckert,
Phys. Rev. Lett. {\bf 101}, 140601 (2008).
\bibitem{Doyon} B. Doyon, Phys. Rev. Lett. {\bf 99}, 076806 (2007).
\bibitem{BVZ} L. Borda, K. Vlad\'ar, and A. Zawadowski,
Phys. Rev. B {\bf 75}, 125107 (2007).
\bibitem{Datta} see e.g. S. Datta, {\em Electronic transport in
mesoscopic systems} (Cambridge University Press, 1997)
\end{thebibliography}
\end{document}